\documentclass[sort&compress]{elsarticle}
\usepackage[bookmarks=false,urlcolor=blue]{hyperref}
\usepackage{amsmath}
\usepackage{amsfonts}
\usepackage{amssymb}
\usepackage{bbold}
\usepackage{framed}
\usepackage{xfrac}
\usepackage{graphicx}
\usepackage{gensymb}
\usepackage{booktabs}
\usepackage{multirow}
\hypersetup{urlcolor=blue}
\usepackage[all]{hypcap}
\usepackage{color}

\begin{document}

\title{Fabrication and characterization of a lithium-glass-based composite neutron detector}


\author[llnl,unc,tunl]{G.C. Rich\corref{cor}}
\ead{grayson@tunl.duke.edu}
\cortext[cor]{Corresponding author}

\author[llnl]{K. Kazkaz}
\ead{kareem@llnl.gov}

\author[llnl]{H.P. Martinez}

\author[llnl,sfsu]{T. Gushue\fnref{fn1}}
\fntext[fn1]{Present address: Twitter, Inc., San Francisco, CA 94103, United States}


\address[llnl]{Lawrence Livermore National Laboratory, Livermore, CA 94550, United States}

\address[sfsu]{Department of Physics and Astronomy, San Francisco State University, San Francisco, CA 94132, United States}

\address[unc]{Department of Physics and Astronomy, University of North Carolina at Chapel Hill, Chapel Hill, NC 27599, United States}

\address[tunl]{Triangle Universities Nuclear Laboratory, Durham, NC 27708, United States}

\begin{abstract}
A novel composite, scintillating material intended for neutron detection and composed of small (1.5 mm) cubes of KG2-type lithium glass embedded in a matrix of scintillating plastic has been developed in the form of a 2.2 in.-diameter, 3.1 in.-tall cylindrical prototype loaded with $\left( 5.82 \pm 0.02 \right)\%$ lithium glass by mass. 
The response of the material when exposed to ${}^{252}$Cf fission neutrons and various $\gamma$-ray sources has been studied; using the charge-integration method for pulse shape discrimination, good separation between neutron and $\gamma$-ray events is observed and intrinsic efficiencies of $\left( 1.15 \pm 0.16 \right)\times 10^{-2}$ and $\left( 2.28 \pm 0.21 \right)\times 10^{-4}$ for ${}^{252}$Cf fission neutrons and ${}^{60}$Co $\gamma$ rays are obtained; an upper limit for the sensitivity to ${}^{137}$Cs $\gamma$ rays is determined to be $< 3.70 \times 10^{-8}$.
The neutron/$\gamma$ discrimination capabilities are improved in circumstances when a neutron capture signal in the lithium glass can be detected in coincidence with a preceding elastic scattering event in the plastic scintillator; with this coincidence requirement, the intrinsic efficiency of the prototype detector for ${}^{60}$Co $\gamma$ rays is $\left( 2.42 \pm 0.61 \right)\times 10^{-6}$ while its intrinsic efficiency for unmoderated ${}^{252}$Cf fission neutrons is $\left( 4.31 \pm 0.59 \right)\times 10^{-3}$.
Through use of subregion-integration ratios in addition to the coincidence requirement, the efficiency for $\gamma$ rays from ${}^{60}$Co is reduced to $\left( 7.15 \pm 4.10 \right) \times 10^{-7}$ while the ${}^{252}$Cf fission neutron efficiency becomes $\left( 2.78 \pm 0.38 \right) \times 10^{-3}$.
\end{abstract}


\maketitle

\section{The ${}^3$He supply problem}
The well-documented shortage of ${}^3$He \cite{kouzes3HeSupplyProblem} has motivated numerous investigations into novel neutron detector technologies which can suitably replace ${}^3$He detectors in their many applications.
Replacement of ${}^3$He-based detection systems is not trivial, however, as they are robust with a very broad application space.

Any competitive replacement must boast several particularly important characteristics: reasonable neutron detection efficiency across a broad range of neutron energies; limited sensitivity to, or the ability to discriminate against, $\gamma$ rays; and a stable efficiency for neutron detection in a mixed radiation field, where both neutrons and $\gamma$ rays are present.
While boron-lined proportional counting tubes have shown promise \cite{lintereur2009,kouzes2010}, several groups have been working on novel neutron detection materials with promising recent results, notably Cs${}_2$LiYCl${}_6$ (or simply CLYC) \cite{clycPSA} and PSD-enabled plastic scintillators \cite{zaitseva2012,zaitseva2013}.
Other groups have investigated composite scintillators -- heterogenous materials composed of neutron-sensitive grains embedded in a supporting plastic matrix -- for their potential application in both basic neutron detection \cite{kazkaz2013,knoll1987,knoll1988} and capture-gated neutron spectrometry \cite{flaska2008,menaa09}.

\section{Operating principles of composite scintillators}
Neutrons incident on a composite detector are intended to interact predominantly through two mechanisms: scattering on nuclei in the supporting, plastic matrix and capture on nuclei in the embedded grains. 
Though many other design features may vary, the plastic matrix serves as an effective moderator for fast neutrons incident on a composite detector, making such detectors sensitive to a broad range of incident neutron energies without the need for additional moderation \cite{kazkaz2013}. 
Early composite detector work by Knoll \emph{et al.} embedded thin-walled glass spheres containing high-pressure ${}^3$He in scintillating plastic; neutrons would enter the volume and be captured on ${}^{3}$He nuclei, with the escaping reaction products depositing energy in the scintillating matrix and producing detectable signals \cite{knoll1987, knoll1988}.
Numerous recent efforts have focused on composites loaded with the inorganic, neutron-sensitive scintillator lithium gadolinium borate (LGB), which have shown promise as both a neutron detector \cite{czirrCaptureGated, kazkaz2013, flaska2008, menaa09} and an antineutrino detector \cite{nathanielAntinu}; in these composites, the scintillation light produced by neutron capture on any of the constituent nuclei of LGB originates within the embedded grains themselves.

LGB-based composites have been fabricated with both scintillating and non-scintillating matrices, but the use of scintillating plastic can provide a mechanism by which neutron- and $\gamma$-ray-generated signals could be distinguished.
In the case of these composites with scintillating matrices, neither the embedded LGB nor the plastic matrix have inherent pulse-shape discrimination capabilities; while PSD-capable plastic has been developed \cite{zaitseva2012,zaitseva2013}, to our knowledge it has not yet been used as a matrix for composite scintillators.
Despite the lack of inherent PSD in either utilized material, the characteristic decay times of the pulses originating in LGB are distinct from those of pulses originating in the plastic scintillator; by virtue of the fact that composite scintillators are largely composed of plastic and that the cross section for capture of low-energy neutrons on the nuclei in the embedded scintillator is very high, it can be argued that $\gamma$-ray pulses have timing characteristics similar to the plastic matrix while neutron-capture pulses have timing characteristics similar to the embedded scintillator \cite{kazkaz2013}.

With the successes of composites utilizing LGB, there is reason to explore potential areas for improvement.
Though each of the eponymous atomic constituents of LGB have isotopes with large thermal-neutron capture cross sections, only captures on ${}^6$Li result exclusively in charged-particle emission; captures on other isotopes in the LGB can result in $\gamma$-ray emission which may be largely indistinguishable from an external $\gamma$-ray background.
There also exist other scintillators whose index of refraction is better matched to that of plastic scintillator $\left( n = 1.58 \right)$ \cite{eljenWebsite} than LGB $\left( n=1.66 \right)$ \cite{czirrCaptureGated}; this mismatch can adversely affect the scintillation-light collection efficiency for larger volumes or for composites with higher concentrations of embedded grains.

\section{Design and fabrication of a lithium-glass-based prototype composite} \label{sec:designAndFabrication}

The materials used in the prototype, as well as the geometry of the embedded scintillator grains, were chosen in an effort to maximize neutron sensitivity and minimize sensitivity to $\gamma$ rays.
This was informed and motivated by recent experimental and simulation work by Kazkaz \emph{et al.}, who explored the use of numerous materials as an alternative to LGB as the embedded scintillator \cite{kazkaz2013}. 
The composite of Ref. \cite{kazkaz2013} used as its matrix EJ-290, a polyvinyl toluene (PVT)-based plastic scintillator from Eljen Technology \cite{eljenWebsite}, and this same material was selected as the matrix for the present composite.
Numerous materials were considered as candidates to serve as the embedded, neutron-sensitive scintillator.
Consideration was also given to the number density of the element on which most neutron captures would occur and the ability to produce the material enriched in the capture isotope.
To improve upon the neutron efficiency and PSD capabilities realized with LGB-based composites, it was desirable to select a material with an isotope on which neutron capture results exclusively, or predominantly, in the emission of charged particles; a related concern is that the material possess a minimal number of isotopes on which neutrons are likely to capture and produce signals which are difficult to distinguish from $\gamma$-ray backgrounds due to the release of $\gamma$ rays after capture, effectively competing for neutrons with the isotopes which produce charged particles after capture, thereby potentially reducing both the neutron detection efficiency and the neutron/$\gamma$ discrimination capabilities of the composite.
Other important factors included: the index of refraction, which should be closely matched to that of plastic scintillator; the light output; the decay time, which is ideally distinct from that of plastic scintillator; and the quenching factor for recoiling nuclei, which determines the electron-equivalent energy of neutron capture signals in the material. 
An extensive discussion on alternative materials can be found in Ref. \cite{kazkaz2013}.
Ultimately, KG2-type lithium glass was selected for the prototype composite for its high atomic fraction of lithium, its enrichment in ${}^6$Li, its index of refraction, and its desirable scintillation decay time \cite{neutronScintillatingGlassesPt1,neutronScintillatingGlassesPt3}.

Successful fabrication of a composite featuring a continuous, even distribution of the embedded scintillator pieces is difficult due to settling of the pieces and formation of bubbles during curing of the plastic matrix. 
To address the issue of settling, a stratified geometry, where cubes of the scintillator are located only at discrete heights along the axis of the detector, was considered; such a geometry can be fabricated by addition of successive layers of scintillator cubes and uncured EJ-290 on top of previous, semi-cured and nearly-solid layers.
Monte Carlo simulations were carried out using the LUXSim \cite{luxsimPaper} front end for \textsc{geant4} \cite{geant4agnostinelli, geant4allison} to compare the intrinsic neutron detection efficiencies of different potential geometries for both thermal and ${}^{252}$Cf-fission-spectrum neutrons: within statistical uncertainties, there was no distinction between the stratified and continuous distributions of embedded scintillator cubes.
Simulations comparing the neutron detection efficiencies as a function of the number of layers present in the stratified geometry suggested there was little dependence on this parameter.
After a qualitative evaluation balancing the areal density of Li-glass cubes on each layer and the distance between adjacent layers, it was decided that the prototype would be divided into 11 layers.
We ignored the effect of optical photon propagation and absorption when determining the optimal number of layers.

In preparation, a large boule of KG2-type lithium glass purchased from Applied Scintillation Technologies \cite{ASTwebsite} was diced into $1.5$-mm cubes; these cubes were not polished, though specimens with obvious damage from the machining process were rejected.
The total mass of lithium glass added to the sample was 12.14 $\pm$ 0.03 g, divided equally among the 11 layers.
The prototype detector was fabricated at Lawrence Livermore National Laboratory (LLNL) over the course of 10 days and was carried out in a glass vessel using an incremental, additive approach. 
EJ-290 resin, with catalyzing agent added in proportions prescribed by Eljen \cite{ej290instructions}, was added to the vessel and partially polymerized by heating through submersion in an oil bath of temperatures ranging between 54 and $60~\degree\text{C}$ for between 6 and 12 h.
Cubes of lithium glass were then added on top of this partially-cured layer, distributed as evenly as possible across the surface.
The polymerization of the underlying layer was sufficient to increase the viscosity to a point where the added cubes remained largely on top of, or very near, the surface; following addition of the glass cubes, another layer of EJ-290 resin was added to the vessel. 
With the resin added, the fabrication vessel was placed in a desiccator which was subsequently evacuated using a small diaphragm pump. 
The rough vacuum in the desiccator removed much of the air trapped in the scintillator resin and effectively prevented the permanent formation of bubbles in the prototype as curing took place.
Following this evacuation procedure, the top-most scintillator layer was partially polymerized and the steps described here were repeated until the desired number of layers was reached.
Throughout the fabrication procedure, dry-nitrogen gas was flowed over the prototype vessel when possible to minimize oxidation of the plastic scintillator. 

\begin{figure*}[]
	\centering
	\includegraphics[width=0.95\textwidth]{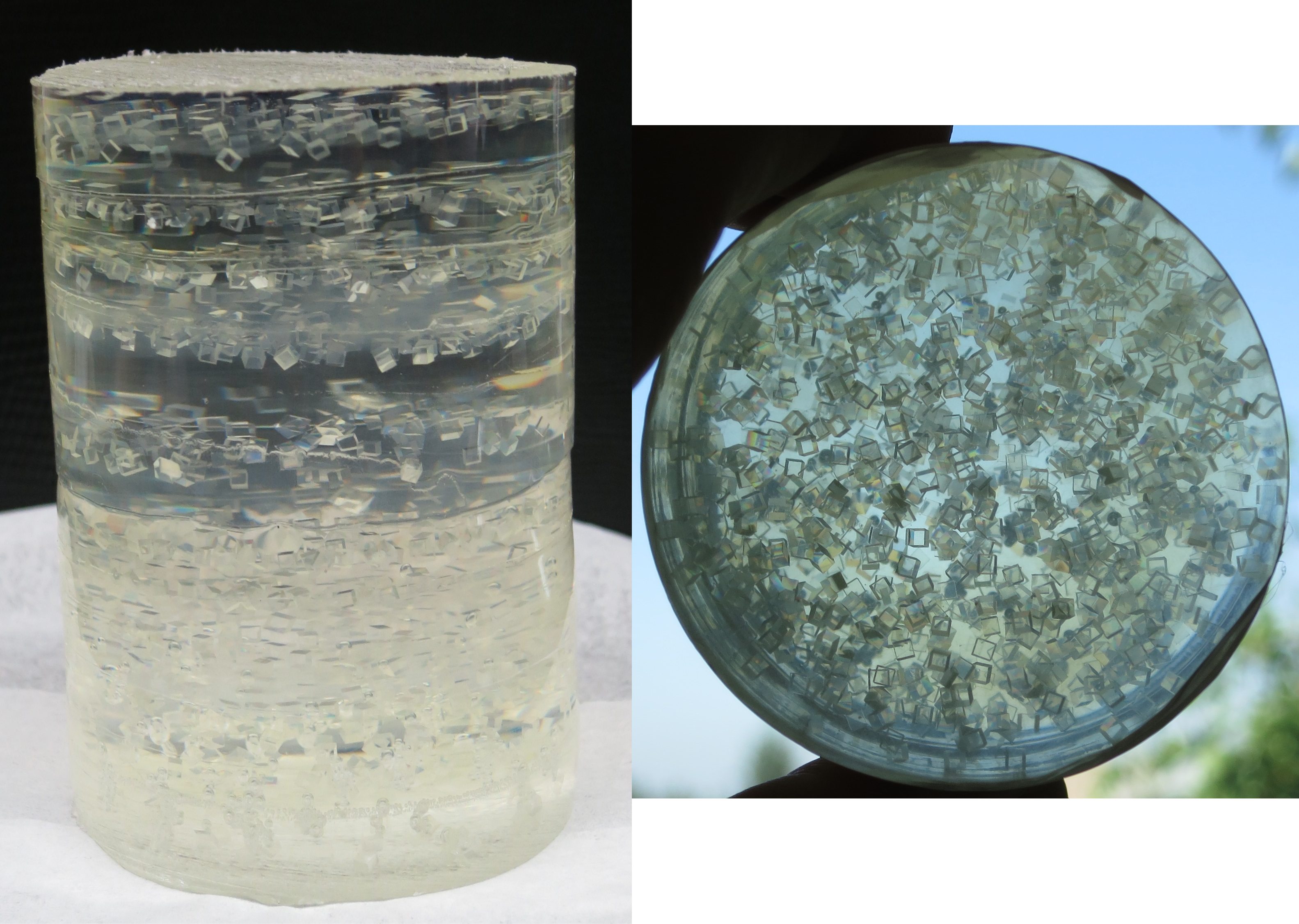}
	\caption[Images of prototype detector.] {\label{fig:viewOfPrototype} Images of the prototype composite scintillator. (Left) Side view of the scintillator prior to final preparation of the circular faces. The stratified nature of the geometry is visible. (Right) Axial view of the scintillator after final preparation of the faces and wrapping in reflective teflon tape. Decent transmission of California sunshine through the prototype scintillator volume is apparent.}
\end{figure*}

After the final layer had been added and evacuated, the vessel was moved to a convection oven for final curing at $\sim\!70~\degree\text{C}$ for approximately 4 days.
The vessel was then removed from the oven and allowed to cool for approximately an hour in a $\sim\!40~\degree\text{C}$ oil bath.
Upon removal from the bath, the glass was scored and the sample placed in a freezer for 2 h before the glass was carefully broken and the composite removed.
The inner diameter of the fabrication vessel was not completely uniform and varied slightly, and this non-uniformity is also present in the detector.
After shaping and polishing, the prototype was found to have a diameter of 2.2 $\pm$ 0.1 in., a height of 3.1 $\pm$ 0.2 in., and a total mass of 208.78 $\pm$ 0.09 g with a lithium-glass mass fraction of $\left(5.82 \pm 0.02\right)$\%.
The prototype composite scintillator can be seen in Fig. \ref{fig:viewOfPrototype}.

\section{Characterization}
Following the fabrication of the prototype, its performance was evaluated.
Detection efficiencies were determined using radioactive sources: ${}^{60}$Co and ${}^{137}$Cs were used to evaluate the efficiency for detection of $\gamma$ rays and a ${}^{252}$Cf source was used to evaluate the efficiency for detection of neutrons in a mixed-field environment, as the spontaneous fission of ${}^{252}$Cf produces both fission neutrons and $\gamma$ rays.
For both the ${}^{252}$Cf and ${}^{60}$Co sources, 24 hours of data were taken; the ${}^{137}$Cs dataset corresponds to a run of 19.8 hours in duration.
Each source was placed in the same position at 22.48 $\pm$ 0.52 in. from the central axis of the detector.
The activities of the sources at the time of the data taking were 6.979 $\mu$Ci for the ${}^{60}$Co source, 9.409 $\mu$Ci for ${}^{137}$Cs, and 1.35 $\mu$Ci $\pm 10\%$ for the ${}^{252}$Cf source; the uncertainties associated with the activities of the $\gamma$-ray sources are negligible.
A 36-hour data set was taken with no radioactive sources present to provide a background spectrum.
During data collection, the composite was located approximately 65 $\pm$ 0.25 in. above a concrete floor, supported by a thin ($\sim0.130$ in.) aluminum surface.

A pair of 2''-diameter photomultiplier tubes (PMTs) was mounted to the prototype detector, one tube on each circular face.
The total observed scintillation light of an event in the scintillator $L_\text{obs.}$ can be reconstructed from the light observed by the two PMTs, $L_1$ and $L_2$, using the prescription of Ref. \cite{nathanielAntinu}, where an effective attenuation length $\alpha_\textrm{eff}$ is assumed; this leads to the expression
\begin{align}
	L_\text{obs.} &=  \sqrt{ \left( L e^{-x/ \alpha_\textrm{eff}} \right) \left( L e^{x/\alpha_\textrm{eff} }  \right) e^{-D/\alpha_\textrm{eff}} } \nonumber \\
	&= \sqrt{L_1 L_2 e^{-D/\alpha_\textrm{eff}}} \nonumber \\
	L_\text{obs.} &\propto \sqrt{L_1 L_2}, \label{eq:Ereconstruct}
\end{align}
where $x$ is the location along the axis of the detector at which the interaction took place and $D$ is the total length of the detector.
In the following discussion, proportional response of the PMTs is assumed and the observed light yield $L_\text{obs.}$ and the energy of an event $E$ are treated as equivalent with some proportionality constant.
The signals from the PMTs were digitized by a Struck 3320-250 digitizer, using the ``n/gamma" firmware available from Struck \cite{struckFirmware}, sampling at 200 MHz.
The digitizer firmware recorded the integrals of 8 ``accumulator" regions: these are user-definable sample ranges which can be used in a variety of ways, some of which are discussed in Sections \ref{section:basicPSD} and \ref{section:gateRatioPSD} as well as in the work by Kazkaz \emph{et al.} \cite{kazkaz2013}.
The present analysis relied exclusively on data from these accumulators; the configuration of the accumulator regions used here are shown in Fig. \ref{fig:PSD_demo}, alongside some example waveforms showing the relative timing of the accumulators with respect to physics events.
The total event size in memory is dramatically reduced when the waveform is not saved, so the reliance on these accumulators for this analysis suggests that it can be utilized with a similar data acquisition system (DAQ) in a high-rate environment: Kazkaz \emph{et al.} estimate that the accumulator-based use of this DAQ is viable up to event rates of several hundred kilohertz while a similar DAQ with waveform readout enabled would be limited to tens of kilohertz \cite{kazkaz2013}.

\subsection{Pulse shape discrimination via the charge-integration method}\label{section:basicPSD}

\begin{figure*}[]
	\centering
	\includegraphics[width=0.95\textwidth]{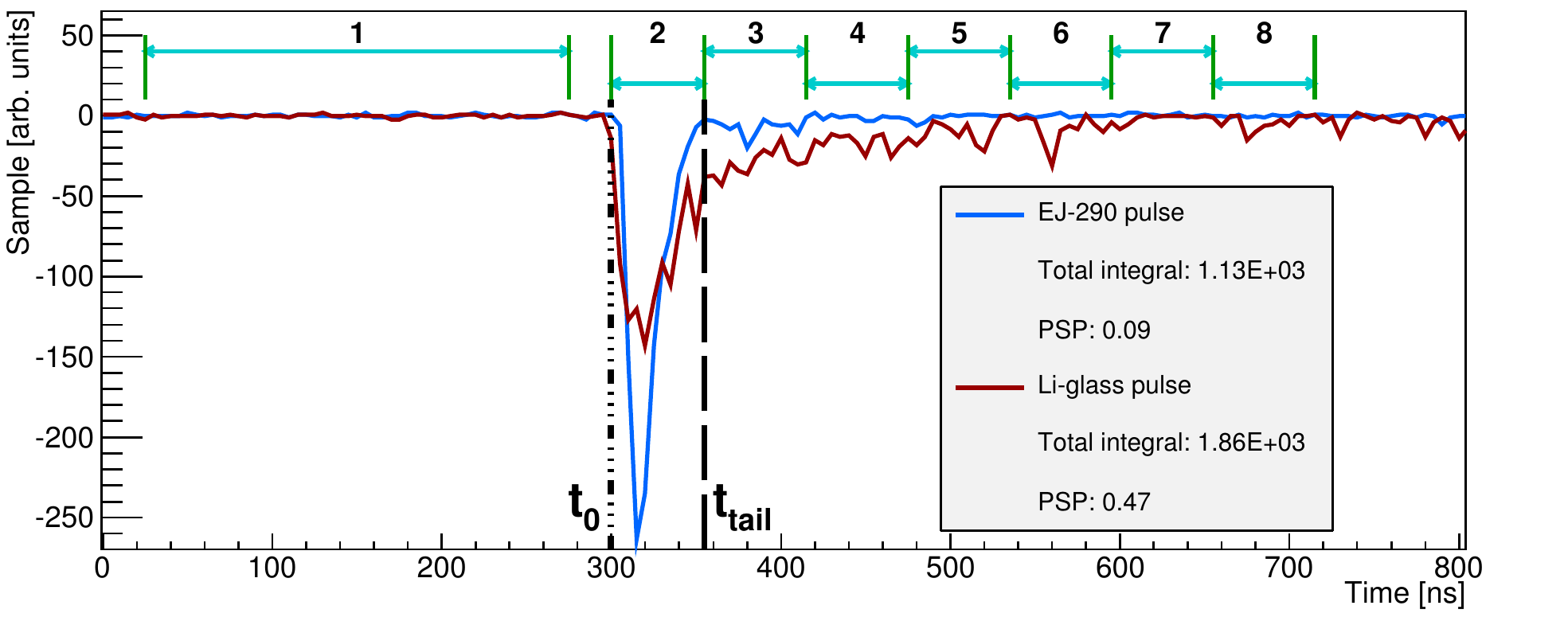}
	\caption[Demonstration of PSD.]{\label{fig:PSD_demo} Example pulses from the prototype composite demonstrating the charge integration method of PSD used in this analysis. These pulses were recorded by the digitizer connected to one of the two PMTs mounted to the detector; as such, these pulses do not represent those which have undergone energy reconstruction using Equation \protect{\eqref{eq:Ereconstruct}}, but they nonetheless serve as a demonstration of the character of the pulses in the detector. The times corresponding to the start of the pulse and to the start of the ``tail" region of the pulse are indicated and labeled $t_0$ and $t_\textrm{tail}$, respectively. The blue pulse is a low-PSP event and would be attributed to an interaction which took place in the EJ-290 matrix. The red pulse is a high-PSP event and would be attributed to an interaction inside one of the Li-glass pieces, likely a neutron capture on ${}^6$Li. Though the pulses have comparable total integrals, they are easily distinguishable by their pulse shape parameters. The configuration of the eight accumulator regions, discussed in Section \protect{\ref{section:gateRatioPSD}}, is also shown. (For interpretation of the references to color in this figure caption, the reader is referred to the web version of this paper.)}
\end{figure*}

The charge-integration method of pulse shape discrimination, more informally known as the  ``tail-to-full ratio" method, determines a pulse shape parameter (PSP) value for each pulse,
\begin{equation} \label{eq:PSP}
	\text{PSP} \equiv \frac{\int_{t_\textrm{tail}}^{t_f} \! Q\left(t\right) dt}{\int_{t_0}^{t_f} \! Q\left(t\right) dt},
\end{equation}
where $Q\left(t\right)$ is the reconstructed signal from the PMTs at time $t$ through use of Eq. \eqref{eq:Ereconstruct}, and times $t_0$, $t_f$, and $t_\textrm{tail}$ correspond to the start of the pulse, the end of the pulse, and the time at which the ``tail" of the pulse begins, respectively.
Example pulses from the composite detector which can be attributed to events in the different materials and show very different PSPs despite similar total charge can be seen in Fig. \ref{fig:PSD_demo}.

To calculate the PSP described by Eq. \eqref{eq:PSP}, the sum of the integrals for accumulators 3-8 was used to represent the charge in the ``tail" of the pulse while the sum of the integrals for accumulators 2-8 was used as the charge in the ``full" pulse.
The integral in accumulator region 1 was used, after scaling by the ratio of the number of samples included in each region, to perform baseline subtractions for each accumulator. 

\begin{figure*}[]
	\centering
	\includegraphics[width=0.95\textwidth]{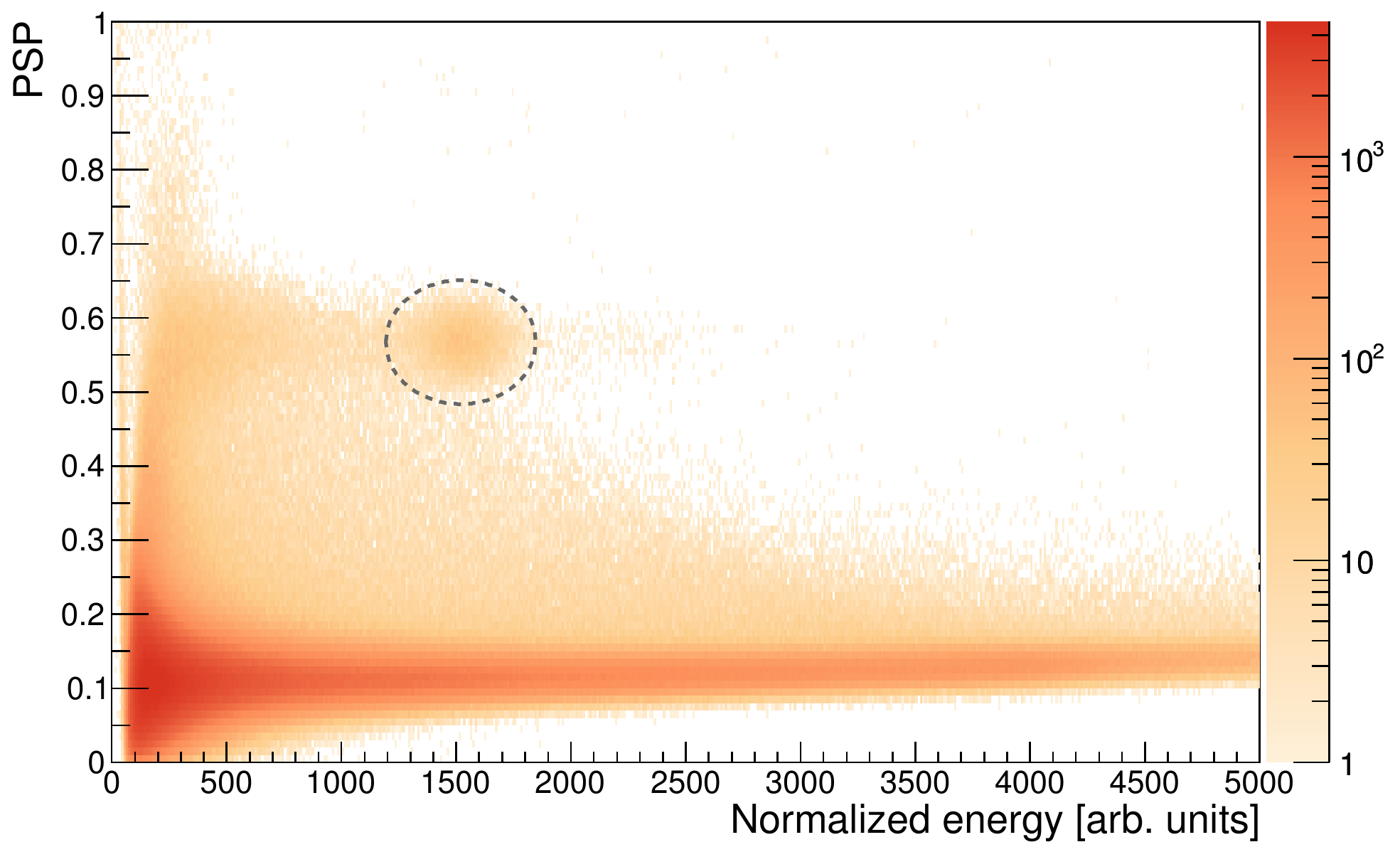}
	\caption[Background 2-D PSD plot.]{\label{fig:backgroundPSD} Events collected in the prototype detector during a 36-hour period with no radioactive sources present. The $x$ axis is the total energy of an event as determined by Eq. \protect{\eqref{eq:Ereconstruct}}, multiplied by a normalization factor discussed in the text; the background data shown here was used to determine the normalization factor for other data sets, and is itself unchanged by normalization. The $y$ axis is the pulse shape parameter (PSP) for an event, or ``tail-to-full ratio": defined quantitatively in Eq. \protect{\eqref{eq:PSP}}, the PSP corresponds to the ratio of charge collected in the ``tail" of an event, after time $t_\text{tail} = 55$ ns, to the total charge collected. Several features can be identified: the band with PSP $\sim$0.1 results from events within the EJ-290 plastic scintillator; the band at PSP $\sim$0.55  can be attributed to events within the lithium glass; and the island at PSP $\sim$0.55 and energy $\sim$1550 corresponds to neutron capture events in the glass. In this 36-hour period, $\left(8.80 \pm 0.09 \left(\text{stat.}\right)\right)\times 10^{3}$ counts were recorded in the 2-dimensional neutron region-of-interest (ROI) indicated by a dark-gray, dashed line. The ROI is defined by the $3\sigma$ level surface of a 2-dimensional Gaussian fit to the neutron-capture peak in this data.}
\end{figure*}

Each pulse was added to a 2-dimensional PSD histogram, with its $x$ and $y$ positions determined by its total integral (Energy) and its PSP, respectively.
To mitigate the effects of any potential drift from either the PMT electronics or temperature dependence in the scintillators between acquisition periods, the $x$ axes are normalized by a factor described later.
An example of the resulting PSD histogram for the 36-hour background data set is shown in Fig. \ref{fig:backgroundPSD}; several key features can be identified: the band of events with PSP $\sim$0.1 correspond to interactions within the EJ-290 plastic scintillator matrix, while the band of events at PSP $\sim$0.55 correspond to interactions within the embedded lithium-glass cubes.
Finally, within the PSP band corresponding to lithium-glass events, the island centered around normalized energy $\sim$1550 is attributed to neutron capture events on the ${}^6$Li content of the lithium glass.

To ensure a consistent location of the neutron-capture region of interest (ROI) between different data sets, a normalization of the $x$ axis was carried out.
To determine the normalization factors $f_\text{N}$, the region defined by $0.49 \leq \text{PSP} \leq 0.65$ for each raw data set (i.e., prior to background subtraction) was projected onto the $x$ axis.
For each data set, a Gaussian was then fit to the peak centered at energy $\sim$1550, corresponding to the neutron capture ``island" visible in Fig. \ref{fig:backgroundPSD}, and the mean parameters from these Gaussian fits were recorded, $b_\text{Cf}, b_\text{Cs}, b_\text{Co}, b_\text{b.g.}$.
The background data set was then used as the standard, so that the normalization factor $f_\text{N}$ applied to each point in, for instance, the ${}^{60}$Co data set was determined by $f_\text{N,Co} = b_\text{Co} / b_\text{b.g.}$.
The determined normalization factors were all within 3\% of 1.0.
Dead time for each run was determined by examination of the difference between timestamps of sequential digitized events and used to determine a scaling factor for the collected data; the dead time fraction for the ${}^{137}$Cs data set was 1.53\% and $< 1\%$ for all other data sets.
Following the energy calibration and dead-time correction, the background was subtracted bin-by-bin after normalizing the bin contents by the ratio of the radioactive-source data collection run time and the background run time.

\begin{figure*}[]
	\centering
	\includegraphics[width=\textwidth]{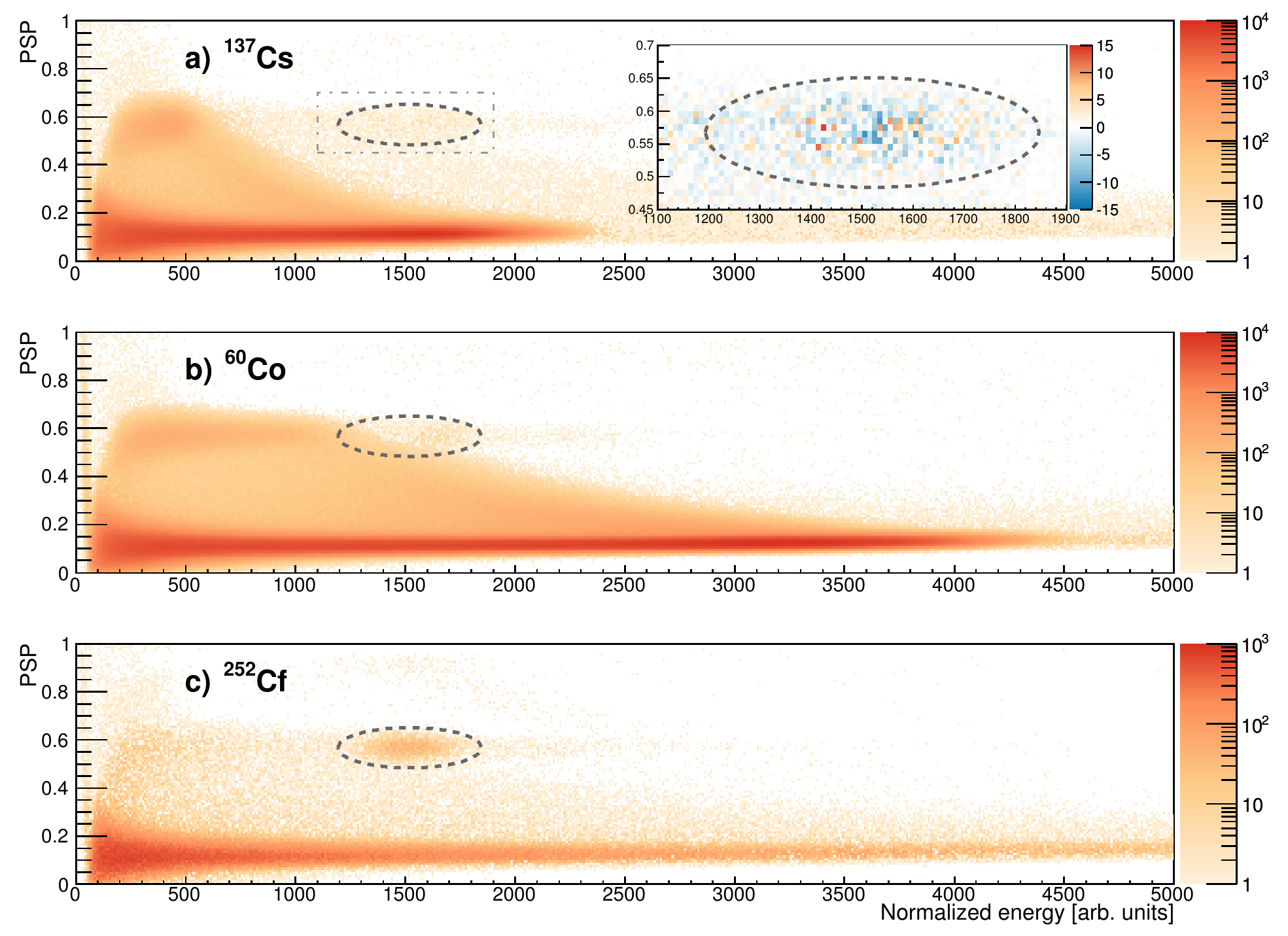}
	\caption[Background subtracted PSD plots.]{\label{fig:backgroundSubtractedPSDplots} Background-subtracted, energy-normalized data collected using unmoderated sources: a) ${}^{137}$Cs; b) ${}^{60}$Co; and c) ${}^{252}$Cf. The region of interest (ROI) in which neutron capture events are located is indicated by a dashed line and is defined as the $3\sigma$ level surface of a 2-dimensional Gaussian fit to the neutron-capture region of the background data set. The intrinsic efficiency, defined as the integral of the ROIs divided by the number of incident $\gamma$ rays or neutrons, for each source is: $< 3.70 \times 10^{-8}$ for ${}^{137}$Cs; $\left( 2.28 \pm 0.21 \right) \times 10^{-4}$ for ${}^{60}$Co; and $\left( 1.15 \pm 0.16 \right) \times 10^{-2}$ for ${}^{252}$Cf. As a consequence of the logarithmic scale of the $z$ axis, bins with contents $\leq$0 resulting from bin-by-bin background subtraction can not be clearly indicated: for the ${}^{137}$Cs data, the inset histogram shows, with a linear $z$ axis, the region around the ROI indicated by a dash-dotted line.}
\end{figure*}

To establish a region of interest (ROI) for neutron capture events, the $3\sigma$ level surface of a 2-dimensional Gaussian fit to the neutron-capture ``island" in the background data set was used; the fit region was defined by $1400 \leq \text{Energy} \leq 1700$ and $0.49 \leq \text{PSP} \leq 0.65$.
The background-subtracted, normalized data taken using ${}^{137}$Cs, ${}^{60}$Co, and ${}^{252}$Cf sources are shown in Fig. \ref{fig:backgroundSubtractedPSDplots} with the ROI overlaid.
The integral within the ROI of these data sets, divided by the number of incident radiation quanta (neutrons in the case of ${}^{252}$Cf and $\gamma$ rays in the cases of ${}^{60}$Co and ${}^{137}$Cs), provides the intrinsic efficiency of the prototype detector for the respective source. 
These efficiencies are listed in Tab. \ref{tab:psdResults}.
\begin{table*}[htbp]
	\centering
	\begin{tabular}{c c}
		\noalign{\smallskip}
		\toprule
		Source & Intrinsic efficiency\\
		\midrule
		${}^{252}$Cf & $\left( 1.15 \pm 0.16 \right)\times 10^{-2}$ \\
		${}^{137}$Cs & $<  3.70 \times 10^{-8}$ \\
		${}^{60}$Co & $\left( 2.28 \pm 0.21 \right)\times 10^{-4 } $\\
		\bottomrule
	\end{tabular}
	\caption[Summary of PSD results.]{\label{tab:psdResults}Detector intrinsic efficiency for $\gamma$ rays and unmoderated fission neutrons using tail/full PSD. Less than 1 count remained in the ROI following background subtraction for the ${}^{137}$Cs dataset, so an upper limit of the efficiency of the composite scintillator for these gamma rays is reported. }
\end{table*}

\subsection{Enhanced PSD using subregion-integration ratios}\label{section:gateRatioPSD}

The charge-integration method of PSD described in Sec. \ref{section:basicPSD} can be easily implemented in analog DAQ systems utilizing charge-to-digital converters and, as can been seen in the results of Tab. \ref{tab:psdResults}, is moderately effective at discriminating between pulses with different decay time characteristics. 
The use of a digitizer for data acquisition affords greater flexibility; by forming different combinations of the accumulator region integrals, whose configuration is shown in Fig. \ref{fig:PSD_demo}, it is possible to project an event into a different pulse-shape parameter space in which EJ-290 events and lithium-glass events may be more easily distinguished. 
Kazkaz \emph{et al.} utilized this approach and successfully increased neutron/gamma discrimination capabilities in a similar composite scintillator; for a thorough discussion of their procedure and results, see Ref. \cite{kazkaz2013}.

\begin{figure*}[]
	\centering
	\includegraphics[width=0.95\textwidth]{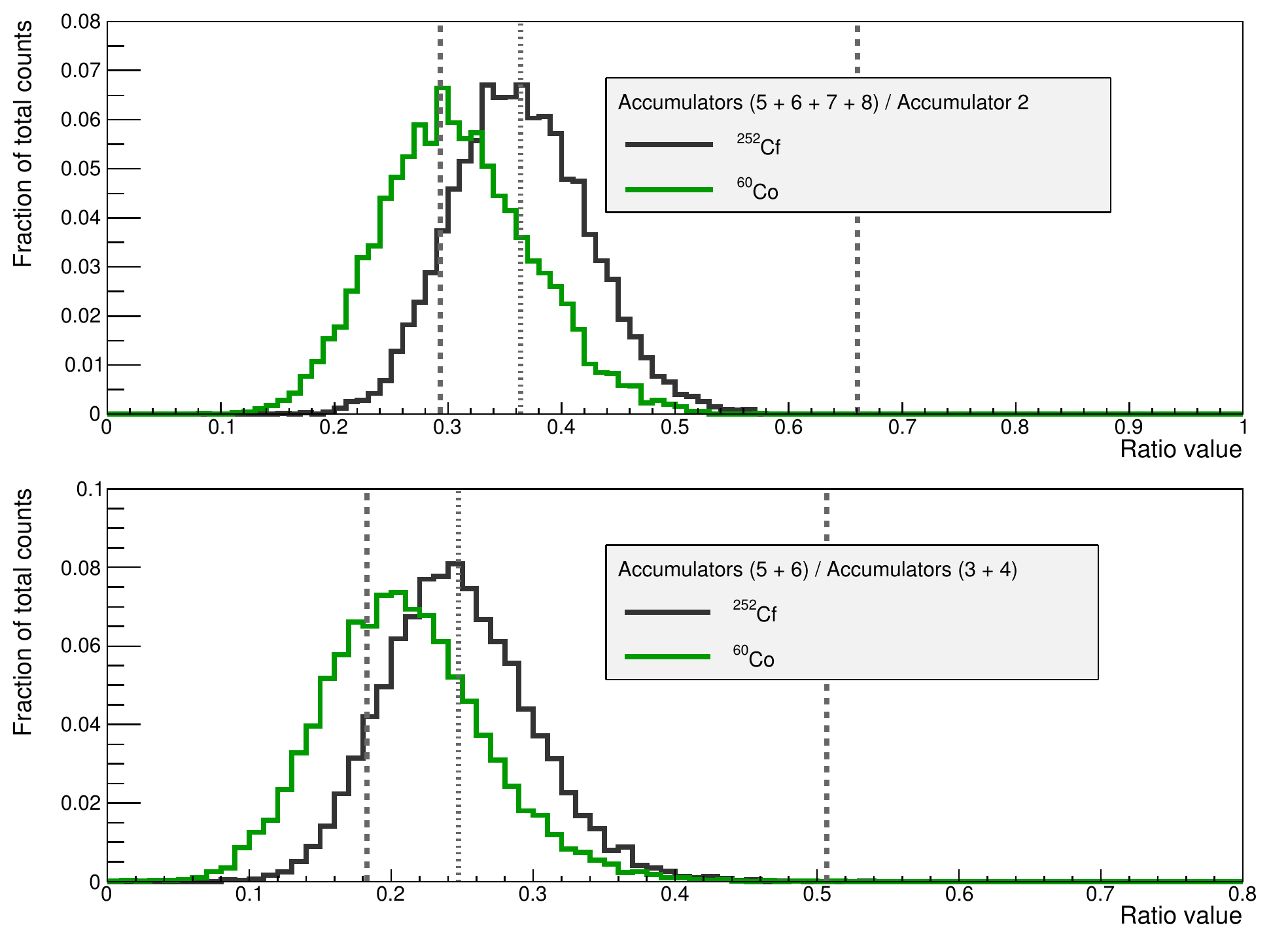}
	\caption[Comparison of accumulator ratios for cobalt and californium events.]{\label{fig:accumulatorRatios} Populations of ${}^{252}$Cf (dark-gray lines) and ${}^{60}$Co (green lines) events for two subregion integral ratios used to enhance pulse-shape-based discrimination of neutron and $\gamma$-ray events. For both combinations of the subregions, or ``accumulators", clear distinctions between the populations are apparent. The top panel shows event populations formed using the sum of accumulators 5, 6, 7, and 8 divided by accumulator 2; the bottom panel shows the data when grouped based on the sum of accumulators 5 and 6 divided by the sum of accumulators 3 and 4. The dotted lines in both panels identify the mean of the ${}^{252}$Cf population, approximated as a Gaussian; the dashed lines represent the boundaries of the cut used to select events, expressed in terms of the mean $\overline{x}$ and $\sigma$ of the Gaussian fit to the ${}^{252}$Cf population in Table \protect{\ref{tab:gateRatioConfigForPSD}}. (For interpretation of the references to color in this figure caption, the reader is referred to the web version of this paper.)}
\end{figure*}

Several different combinations of subregion integrals were considered and their efficacy in discriminating between event types was explored.
To find effective combinations and to tune the cut region for a given combination, histograms of the value of each combination were populated by select events from the ${}^{252}$Cf and ${}^{60}$Co datasets. 
Events selected from the ${}^{60}$Co dataset came from a region of the PSD plot which was near but not overlapping the neutron ROI in energy, $1000 \leq \text{Energy} \leq 1220$, while events selected from the ${}^{252}$Cf dataset were within the neutron ROI, $1365 \leq \text{Energy} \leq 1700$; both populations were required to satisfy $0.52 \leq \text{PSP} \leq 0.61$, safely ensuring that all of the events selected from the ${}^{252}$Cf data are within the neutron ROI.
Examples of ${}^{252}$Cf and ${}^{60}$Co populations for two subregion integral ratios, regions (5 + 6 + 7 + 8)/region 2 and regions (5 + 6)/regions (3 + 4), are shown in Fig. \ref{fig:accumulatorRatios}.

Five combinations of the accumulators were selected for use in this analysis, each of which showed distinction between the ${}^{252}$Cf and ${}^{60}$Co populations.
To define the cut region in each combination, a 1-dimensional Gaussian was fit to the ${}^{252}$Cf population and the region was then defined in terms of the mean $\overline{x}$ and $\sigma$ parameters.
The subregion configurations used and the cut values determined are collected in Tab. \ref{tab:gateRatioConfigForPSD}.
\begin{table*}[htbp]
	\centering
	\begin{tabular}{c c c}
		\noalign{\smallskip}
		\toprule
		Subregion configuration & Minimum value & Maximum value\\
		\midrule
		2 / Total & $\overline{x} - 5 \sigma$ & $\overline{x} + 1.3 \sigma$ \\
		2 / (3 + 4) & $\overline{x} - 1.65 \sigma$ & $\overline{x} + 1.6 \sigma$ \\
		(4 + 5) / (2 + 3) & $\overline{x} - 1.15 \sigma$ & $\overline{x} + 5 \sigma$ \\
		(5 + 6) / (3 + 4) & $\overline{x} - 1.25 \sigma$ & $\overline{x} + 5\sigma$ \\
		(5 + 6 + 7 + 8) / 2 & $\overline{x} - 1.2 \sigma$ & $\overline{x} + 5 \sigma$ \\
		\bottomrule
	\end{tabular}
	\caption[Subregion configuration for PSD.]{\label{tab:gateRatioConfigForPSD}The subregion configurations used to perform PSD and corresponding cut ranges, described in terms of the mean $\overline{x}$ and $\sigma$ parameters of a 1-dimensional Gaussian fit to the population of events generated from the ${}^{252}$Cf data set. In the case of the first combination, ``total" refers to the sum of the integrals from subregions 2 through 8; this combination is similar to the PSP defined in Equation \protect{\eqref{eq:PSP}}. The integral from each subregion is baseline-subtracted using subregion 1. The configuration of the subregions with respect to a waveform can be seen in Figure \protect{\ref{fig:PSD_demo}}. The boundaries presented here for each subregion combination preserve $\sim$90\% of the counts in the ${}^{252}$Cf data when applied individually. These subregion combinations do not necessarily represent unique or orthogonal projections into pulse-shape-parameter space.}
\end{table*}
Though the populations were not strictly normally distributed, the approximation with a Gaussian allowed a more systematic approach to definition of the cut regions.
The boundaries of the cut region for each individual accumulator combination were adjusted by changing the multiplicative coefficient of the $\sigma$ parameter until $\sim$90\% of the ${}^{252}$Cf events would remain, excluding the effects of the other accumulator combination cuts; this is a similar approach to that adopted in Ref. \cite{kazkaz2013}. 
In cases where a higher- or lower-boundary of the region provided no neutron/gamma discrimination power, as can be seen in Fig. \ref{fig:accumulatorRatios} where reducing the upper boundaries of the cut region would preferentially reject ${}^{252}$Cf counts, the boundary was set to $\pm5\sigma$, as appropriate. 
The cuts described in Tab. \ref{tab:gateRatioConfigForPSD} were applied to each reconstructed event and 2-dimensional PSD histograms, like those of Fig. \ref{fig:backgroundSubtractedPSDplots}, were populated with surviving events; to determine the efficiency of the composite detector to the test sources with these additional cuts, these histograms are then integrated with the same ROI determined in Sec. \ref{section:basicPSD}.
Though each individual subregion cut presented in Tab. \ref{tab:gateRatioConfigForPSD} preserves $\sim$90\% of the counts in the ROI for ${}^{252}$Cf, the subregion cuts are not necessarily orthogonal.
Consequently, the cumulative reduction of ROI counts from ${}^{252}$Cf following application of more than one of the subregion cuts is not guaranteed to be a multiple of 10\%.
Results for the efficiency measurements using this subregion-integration technique for PSD are collected in Tab. \ref{tab:psdAndRatioResults}, alongside the results from the charge-integration technique for comparison.

\begin{table*}[htbp]
	\centering
	\begin{tabular}{c c c}
		\noalign{\smallskip}
		\toprule
		\multirow{2}{*}{Source} & \multicolumn{2}{c}{Intrinsic efficiency} \\
		& Charge integration & Subregion integration \\
		\midrule
		${}^{252}$Cf & $\left( 1.15 \pm 0.16 \right) \times 10^{-2}$ & $\left( 4.23 \pm 0.56 \right) \times 10^{-3}$ \\
		${}^{60}$Co & $\left( 2.28 \pm 0.21 \right) \times 10^{-4}$ & $\left( 2.09 \pm 0.27 \right) \times 10^{-5}$ \\
		\bottomrule
	\end{tabular}
	\caption[Summary of efficiency results for charge integration and subregion integration PSD results.]{\label{tab:psdAndRatioResults}Efficiency results for both the charge integration and subregion integration techniques of pulse shape discrimination. Five different subregion, or accumulator, combination cuts described in the text and in Table \protect{\ref{tab:gateRatioConfigForPSD}} are used to establish the values shown. When each subregion cut is applied individually, approximately 90\% of the counts in the region-of-interest (ROI) for the ${}^{252}$Cf data are retained. The subregion cuts are not necessarily orthogonal, however, and consequently the cumulative retention of counts in the ROI after application of multiple cuts is not guaranteed to be a multiple of 0.9. For ${}^{137}$Cs data, less than 1 count remains in the ROI after background subtraction for both techniques, and thus an upper limit of the sensitivity to ${}^{137}$Cs $\gamma$ rays is determined to be $3.70 \times 10^{-8}$.}
\end{table*}

\subsection{Analysis including coincidence requirements and additional cuts} \label{sec:coincidenceRequirement}

The operating principles of capture-gated neutron spectrometers \cite{fisher2011, drakeFeldmanHurlbut, czirrCaptureGated, brooksKlein} expose a potential avenue towards increased discrimination of $\gamma$-ray events beyond PSD alone.
Neutrons incident on the composite detector will have a high likelihood of undergoing elastic scattering interactions with nuclei in the supporting matrix prior to either escaping the detector or participating in a capture reaction. 
The series of scattering interactions serves to moderate the energy of incident fast neutrons, increasing the probability of capture within the detector volume, but each scattering reaction also results in a recoiling nucleus; in the case of scattering in the EJ-290 matrix, these recoiling nuclei produce observable scintillation signals.
For energetic neutrons incident on the composite detector, there is consequently a characteristic signature: elastic scattering in the EJ-290 produces short-timescale pulses which are followed several tens of microseconds later by longer-timescale pulses resulting from neutron capture on ${}^6$Li in the lithium glass. 
Capture-gated spectrometry utilizes the elastic scattering signals to reconstruct the incident neutron energy, focusing only on those followed by capture-like scintillation signals.
By enforcing a coincidence requirement and that the sequence of pulses have neutron-like character (a long-timescale capture pulse preceded by a short-timescale elastic scattering pulse), enhanced discrimination of $\gamma$-ray signals can be realized with only moderate reduction in efficiency for detection of energetic neutrons. 

\begin{figure*}[]
	\centering
	\includegraphics[width=0.95\textwidth]{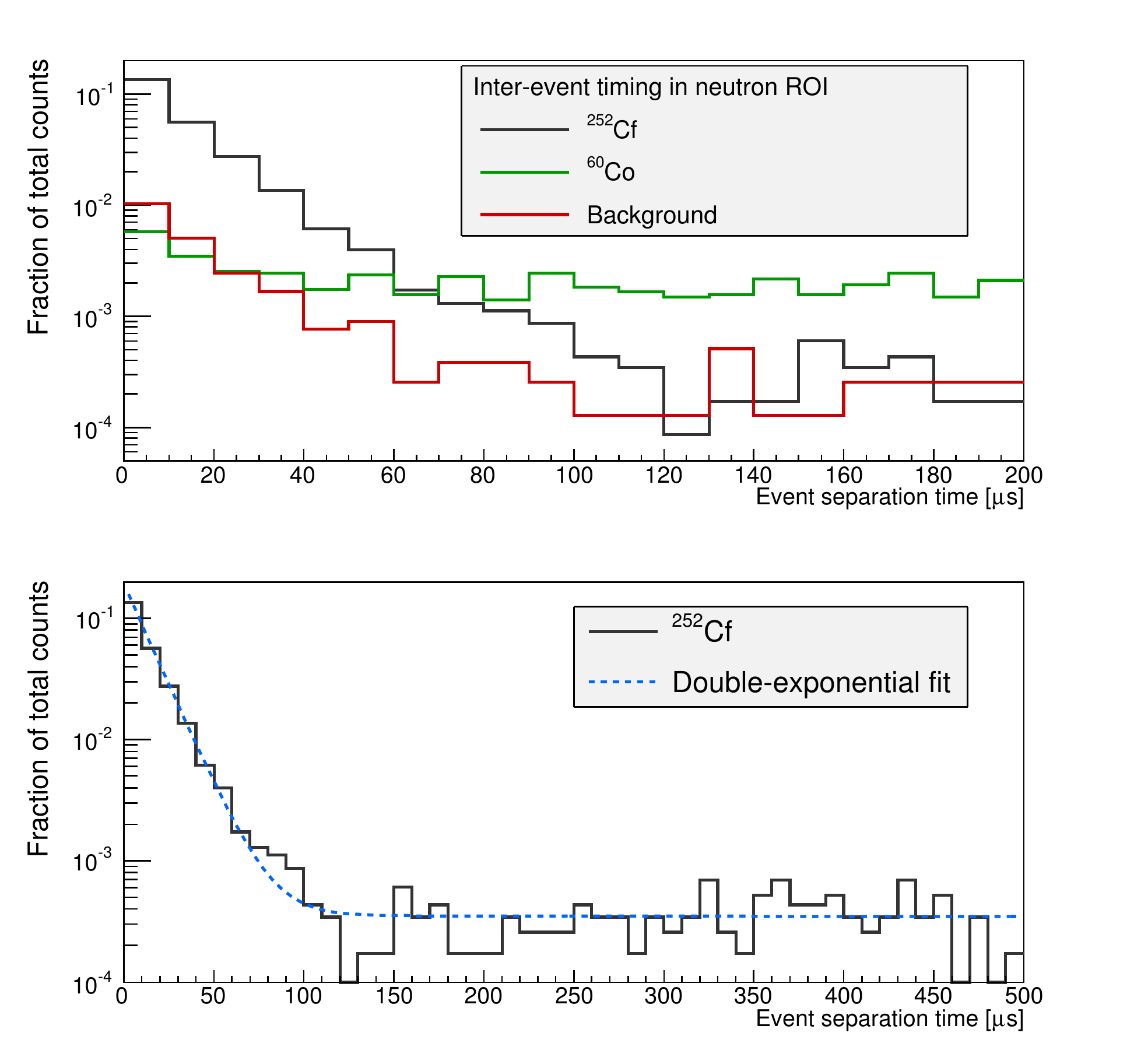}
	\caption[Inter-event separation time in neutron ROI.]{\label{fig:intereventTimingDistribution} (Top) Inter-event timing for pairs of events, wherein the first event satisfies a low-PSP requirement and the second event would populate the neutron ROI, from the ${}^{252}$Cf (dark-gray line), ${}^{60}$Co (green line), and background (red line) data sets. Each data set is normalized such that the full integral of the inter-event timing histogram is equal to 1. (Bottom) The ${}^{252}$Cf inter-event timing population with an overlaid double-exponential fit described in Section \protect{\ref{sec:coincidenceRequirement}} of the text. Assuming the two timing components of the double exponential correspond to accidental coincidences and the distribution of neutron-moderation signals followed by neutron-capture signals, the scattering- or moderation-to-capture time constant for the prototype detector for spontaneous fission neutrons from ${}^{252}$Cf is 13.1 $\mu$s. (For interpretation of the references to color in this figure caption, the reader is referred to the web version of this paper.)}
\end{figure*}

\begin{figure*}[]
	\centering
	\includegraphics[width=0.95\textwidth]{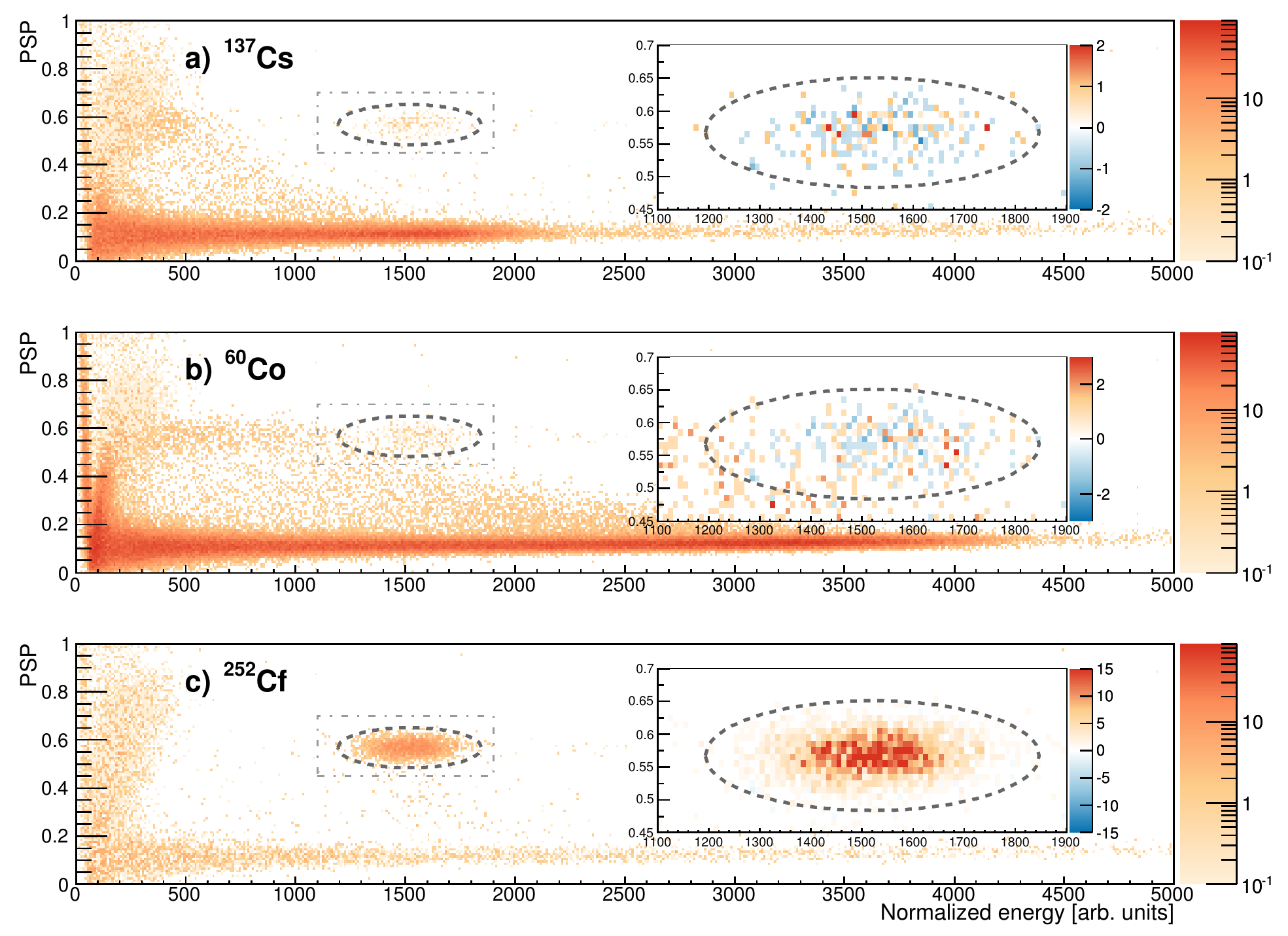}
	\caption[PSD plots with coincidence requirement.]{\label{fig:coincidenceReqPSDplots} Energy-normalized data sets after application of the coincidence requirement and background subtraction for: a) ${}^{137}$Cs; b) ${}^{60}$Co; and c) ${}^{252}$Cf. The events populating these histograms must have occurred within 40 $\mu$s of a preceding event with PSP $< 0.45$; the background data which has been subtracted from the data sets shown here is subjected to the same requirement. The 2-D region of interest (ROI) for neutron capture is indicated by a dashed line. The inset figures show in greater detail the region around the ROI, which is identified by a dash-dotted line in the figures. The effect of the coincidence requirement can be seen by comparing this data with that shown in Figure \protect{\ref{fig:backgroundSubtractedPSDplots}}, noting the difference in $z$-axis scales.}
\end{figure*}

Using the timestamp recorded by the digitizer, the separation in time of sequential events was determined and a coincidence requirement was applied: this cut accepted events only if they occurred within 40 $\mu$s of a preceding event and only if the preceding event fit the character of an event originating in the plastic scintillator (PSP $<$ 0.45).
The duration of the coincidence window was established by examining the relative timing between sequential events in cases where the first of the events satisfied a low-PSP requirement (suggesting the pulse originated in the EJ-290) and the second event would populate the neutron ROI; a plot of these distributions for the ${}^{252}$Cf and ${}^{60}$Co sources, as well as the background data set, is shown in the top panel of Fig. \ref{fig:intereventTimingDistribution}.
A double-exponential distribution of the form $T\left( t \right) = A_1 \exp{ \left( -t / \tau_1 \right) } + A_2 \exp{ \left( -t / \tau_2\right) }$ was used to model the ${}^{252}$Cf data; in this model, the two distributions correspond to accidental coincidences and event pairs generated by neutron elastic scattering in the matrix followed by neutron capture in the lithium glass.
Though the data may contain additional timing structure owing to systematics associated with the source, geometry, data acquisition system, or other origins, the double-exponential approximation is sufficient over a small range of inter-event separation times (0-500 $\mu$s) to extract the scattering-to-capture time constant used to define the coincidence window.
The time constant for accidental coincidences was determined by fitting a single exponential to the ${}^{252}$Cf inter-event timing distribution between separation times of 1000 $\mu$s and 10,000 $\mu$s.
With this time constant fixed, the inter-event timing distribution was fit with the double-exponential function over a range of 0 $\mu$s to 500 $\mu$s; the parameters of this fit for the ${}^{252}$Cf data set were $\tau_1 = 13.1$ $\mu$s, $A_1 = 1.90 \times 10^{-1}$ for the exponential corresponding to the scattering-to-capture time and $\tau_2 = 2.68 \times 10^{4}$ $\mu$s, $A_2 = 3.53 \times 10^{-4}$ for the accidental coincidences. 
The coincidence window of 40 $\mu$s corresponds to acceptance of $\sim$95\% of the event pairs contained within the prompt distribution ($\tau_1 = 13.1$ $\mu$s) established here.
Applying these requirements to all data sets and then performing background subtraction results in the data depicted in Fig. \ref{fig:coincidenceReqPSDplots}, which can be compared to the data shown in Fig. \ref{fig:backgroundSubtractedPSDplots} where the coincidence requirement is not enforced. 

The coincidence requirement in conjunction with the charge-integration technique modestly reduced the ${}^{252}$Cf fission neutron intrinsic efficiency to $\left( 4.31 \pm 0.59 \right) \times 10^{-3}$ while dramatically affecting the intrinsic efficiency for ${}^{60}$Co $\gamma$ rays, which was reduced to $\left( 2.42 \pm 0.61 \right) \times 10^{-6}$.
The subregion-integration technique, combined with the coincidence requirement, yields efficiencies for ${}^{252}$Cf and ${}^{60}$Co of  $\left( 2.78 \pm 0.38 \right) \times 10^{-3}$ and $\left( 7.15 \pm 4.10 \right) \times 10^{-7}$, respectively.
Efficiency for ${}^{137}$Cs $\gamma$ rays remains an upper limit of $< 3.70 \times 10^{-8}$.
It should be noted that by using the coincidence requirement in conjunction with the subregion-integration approach the sensitivity to $\gamma$ rays from ${}^{60}$Co is subject to large uncertainties ($> 50\%$) due to limited statistics.

\begin{figure*}[]
	\centering
	\includegraphics[width=0.95\textwidth]{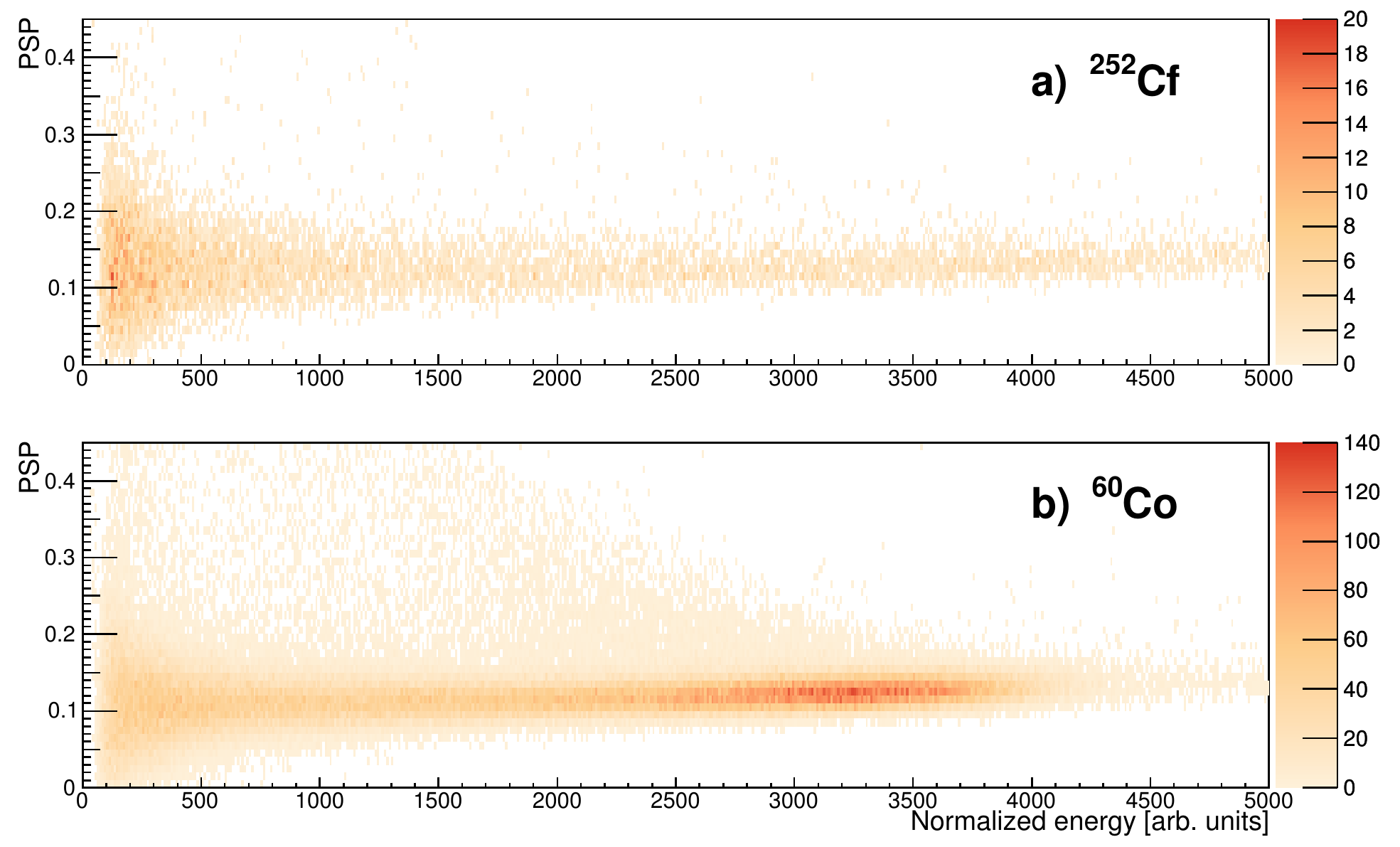}
	\caption[PSD plots of preceding pulses in coincident pairs.]{\label{fig:precedingPulsePSD} PSD plots populated by the initial pulses in coincident pairs for: a) ${}^{252}$Cf and b) ${}^{60}$Co. The concentration of events between energies of $\sim$2000 and $\sim$4000 evident in the ${}^{60}$Co $\gamma$-ray data is not present in the fission neutron data; the energies of these populations are in accord with the energies of the low-PSP single-event data for ${}^{60}$Co and ${}^{252}$Cf shown in panels b) and c), respectively, of Figure \protect{\ref{fig:backgroundSubtractedPSDplots}}. Due to these distinct populations, requiring that the first pulse in a coincident pair have $E \leq 2000$ achieves preferential rejection of $\gamma$-ray interactions.}
\end{figure*}

The $\gamma$-rejection efficacy of the coincidence cut, already very effective, could be slightly improved by noting the energy of the preceding pulse in addition to its PSP: specifically, the ${}^{60}$Co data shows a concentration of counts with $2000 \leq \text{Energy} \leq 4000$, while a similar pattern is not evident in the neutron data collected with a ${}^{252}$Cf source, which yielded predominantly lower-energy preceding pulses; these characteristics can be seen in Fig. \ref{fig:precedingPulsePSD}.
Figure \ref{fig:precedingPulsePSD} shows PSD plots of the \emph{initial} pulses in coincident pairs, of which the \emph{second} pulses populate the data sets shown in Fig. \ref{fig:coincidenceReqPSDplots}, for data taken with a ${}^{252}$Cf source and a ${}^{60}$Co source.
By applying an additional requirement to the coincidence cut -- that the preceding pulse must satisfy $\text{Energy} \leq 2000$ -- $\gamma$-ray rejection was further increased with marginal loss of neutron efficiency; the results of this cut on efficiencies when used with both the charge-integration and subregion-integration methods are shown in Tab. \ref{tab:coincidenceResults}
\begin{table*}[htbp]
	\centering
	\begin{tabular}{p{3cm} c c c}
		\noalign{\smallskip}
		\toprule
		\multirow{2}{*}{Analysis method} & \multirow{2}{*}{Source} & \multicolumn{2}{c}{Intrinsic efficiency}\\
		& & Charge integration & Subregion integration \\
		\midrule
		\multirow{2}{*}{ROI integration} & ${}^{252}$Cf & $\left( 1.15 \pm 0.16 \right)\times 10^{-2}$ & $\left( 7.24 \pm 0.99 \right) \times 10^{-3}$ \\
		& ${}^{60}$Co & $\left( 2.28 \pm 0.21 \right)\times 10^{-4 } $ & $\left( 4.08 \pm 0.49 \right) \times 10^{-5}$ \\
		\midrule
		\multirow{2}{*}{Coincidence} & ${}^{252}$Cf & $\left( 4.31 \pm 0.59 \right)\times 10^{-3}$ &  $\left( 2.78 \pm 0.38 \right) \times 10^{-3}$ \\
		& ${}^{60}$Co & $\left( 2.42 \pm 0.61 \right)\times 10^{-6 } $ & $\left( 7.15 \pm 4.10 \right) \times 10^{-7} $ \\
		\midrule
		\multirow{2}{3cm}{Coincidence with energy cut}  & ${}^{252}$Cf & $\left( 3.88 \pm 0.53 \right)\times 10^{-3}$ & $\left( 2.50 \pm 0.34 \right) \times 10^{-3}$ \\
		 & ${}^{60}$Co & $\left( 1.53 \pm 0.53 \right)\times 10^{-6 }$ & $ \left( 6.66 \pm 3.88 \right) \times 10^{-7}$ \\
		\bottomrule
	\end{tabular}
	\caption[Summary of coincidence requirement results.]{\label{tab:coincidenceResults}Intrinsic detection efficiencies obtained by enforcing a 40-$\mu$s coincidence requirement and by additionally requiring that the total energy of the preceding pulse be $\leq 2000$. For all approaches, background subtraction resulted in less than one count in the ROI for data collected with ${}^{137}$Cs and the upper limit of the sensitivity of the composite to ${}^{137}$Cs $\gamma$ rays is $3.70 \times 10^{-8}$. The results from the analyses of Sections \protect{\ref{section:basicPSD}} and \protect{\ref{section:gateRatioPSD}}, also presented in Tables \protect{\ref{tab:psdResults}} and \protect{\ref{tab:psdAndRatioResults}}, are included here to highlight the dramatic reduction in sensitivity to ${}^{60}$Co $\gamma$ rays through use of the coincidence requirement. It is important to note that the determined sensitivity to $\gamma$ rays from ${}^{60}$Co, in several cases, subject to large statistical uncertainties; more strenuous limits could be placed with longer acquisition periods or stronger sources. }
\end{table*}

\subsection{Comparison with simulation} \label{sec:comparisonWithSims}
A geometry which closely approximated the experimental configuration was developed in MCNP6. 
This model included the concrete floor and several other nearby volumes, including the aluminum surface on which the composite detector was placed.
The material definitions were informed by Ref. \cite{compendium}, with deviations where specific information was available; notably, the density of PVT-based plastic scintillators reported by Eljen Technology \cite{eljenWebsite} (1.023 g/cc) is slightly lower than that suggested in Ref. \cite{compendium} (1.032 g/cc). 
Isotopic abundances were specified where necessary to promote data-table based computation and to ensure the correct 95\% isotopic enrichment of ${}^6$Li in KG2-type lithium glass.

To model a ${}^{252}$Cf spontaneous-fission source, a Watt fission spectrum was used to define the initial energy of the simulated neutrons; the parameters used for this distribution were those recommended in Ref. \cite{mcnp6manual}.
These simulations focused on establishing a benchmark for the neutron-detection efficiency and did not include $\gamma$-ray sources.
The simulated efficiency was determined by dividing the number of captures on ${}^6$Li recorded within the lithium-glass volumes by the number of neutrons incident on the composite as determined by the simulation; the simulation-determined geometric efficiency $\epsilon_\text{G, sim} = 1.20 \times 10^{-3}$, with statistical uncertainty of $1.1 \times 10^{-6}$, was in agreement with an analytical approximation of the geometric efficiency of the composite detector $\epsilon_\text{G} = \left( 1.19 \pm 0.11 \right) \times 10^{-3}$. 
Efficiency effects related to scintillation-light production, photon transport, and light-collection efficiency were not considered.

The MCNP6 simulations carried out predict a detection efficiency for spontaneous-fission neutrons from ${}^{252}$Cf of $\left(1.33 \pm 0.01 \text{(stat.)} \right) \times 10^{-2}$, which is slightly larger than the experimentally-determined efficiency of $\left( 1.15 \pm 0.16 \right) \times 10^{-2}$, obtained when using charge-integration PSD. 
As was observed in Ref. \cite{mayer2015}, simulations indicate that the experimental environment provided a non-negligible contribution to the measured efficiency: a geometry which included only the neutron source, air, and the composite detector yielded an efficiency of $\left( 1.02  \pm 0.01 \left( \text{stat.} \right) \right)\times 10^{-2}$.

\section{Conclusions}

We have fabricated a first-of-its-kind composite scintillator consisting of KG2-type lithium glass cubes embedded in a supporting matrix of scintillating plastic.
The additive, successive approach to fabrication utilized here, where subsequent layers are added after nearly-complete polymerization of the underlying layer, should be extensible to larger volumes and alternative geometries.
This approach yielded a single-volume detector with no persistent optical or mechanical interfaces.
Though our prototype had a non-uniform diameter with some discontinuities between layers, these effects could be mitigated by use of a fabrication vessel with a more regular shape and through more controlled thermal cycling at times when the sample is removed from heating for the purpose of layer addition.

Though the fabrication of a larger-volume sample is conceptually straightforward, simulations should be developed which account for the propagation of scintillation photons through these volumes. 
Through careful mitigation of bubble formation during the fabrication process and through use of lithium glass as the embedded scintillator grain material, which has a closely-matched index of refraction with the PVT-based matrix and whose surfaces were thoroughly whetted by the uncured polymer, we have been able to obtain a volume with good optical transmission properties. 
Despite this success, possible attenuation of scintillation light through larger volumes of a similar composite should be quantitatively considered.

The prototype composite has shown good neutron/$\gamma$ discrimination and neutron detection efficiency, established with ${}^{137}$Cs, ${}^{60}$Co, and ${}^{252}$Cf radioactive sources and summarized in Tab. \ref{tab:psdAndRatioResults}. 
The MCNP6 simulations carried out show decent approximation of the experimental results of neutron detection efficiency from the charge-integration PSD approach for this prototype and suggest that some component of the measured efficiency is due to the experimental geometry. 
Examination of the ${}^{60}$Co data in Fig. \ref{fig:backgroundSubtractedPSDplots} shows that the endpoint of the spectrum from this source in the Li-glass PSP band is close to the ROI associated with neutron capture events in the glass; selection of an alternative embedded scintillator with a higher electron-equivalent signal yield for neutron capture than the KG2-type lithium glass used in this work may offer improved neutron/$\gamma$ discrimination, though the selection of an embedded material involves consideration of many factors (see the discussions in Sec. \ref{sec:designAndFabrication} and Ref. \cite{kazkaz2013}). 
Alternative matrices, such as nonscintillating acrylic, may also yield composite, lithium-glass-based detectors with very effective neutron/$\gamma$ discrimination: Ianakiev \emph{et al.} have shown promising initial results for high neutron/$\gamma$ discrimination efficacy with composites based on GS20-type lithium glass and poly(methyl methacrylate) (PMMA) or mineral oil matrices \cite{ianakiev2014}.

Investigations of the incident-neutron-energy-dependent response and sensitivity of lithium-glass based composites may be fruitful.
As the neutron/$\gamma$-discrimination capabilities of these composites arise from the distinct pulse shapes resulting from $\gamma$-rays and, ultimately, capture of \emph{low-energy} neutrons on lithium nuclei, this class of detectors should yield reliable discrimination between $\gamma$-ray and neutron-induced signals for a broad range of incident neutron energies: irrespective of the incident energy of the neutron, the identifiable signal relied upon for detection results from the release of energetic charged-particles after capture of the neutrons. 
For low-energy incident neutrons ($\lesssim100$ keV), use of the coincidence requirement discussed in Sec. \ref{sec:coincidenceRequirement} may be precluded due to small signal yield from the initial scattering of the neutron in the plastic matrix, though the PSD techniques should remain efficacious for all incident neutrons sufficiently energetic to enter the detector and capture in one of the lithium-glass cubes.
This can be contrasted with the energy-dependent PSD realized in liquid and plastic scintillators, particularly in the latter category where reliable PSD for signals below 200-keVee is challenging \cite{pozziPSD,cester2014}.
In addition to PSD across a broad spectrum of neutron energies, lithium-glass-based composites may also possess the ability to provide information on the incident neutron energy spectrum when operated as capture-gated neutron spectrometers \cite{czirrCaptureGated}: an application space in which LGB-based composites have demonstrated success \cite{menaa09}.

For this initial evaluation of the performance of the prototype composite, no concerted effort was made to optimize the configuration of the accumulator regions of the digitizer for neutron/$\gamma$ discrimination.
The analysis of this work was carried out using exclusively data from these accumulators, and consequently the neutron/$\gamma$ discrimination capability of similar composites could potentially be improved beyond that demonstrated here while still utilizing the subregion integration technique and accumulator-based digitizer readout, and thus remaining feasible for high-event-rate environments where DAQ deadtime would impact performance of systems relying on waveform readout.
Examination of the ${}^{60}$Co events which appear in the neutron ROI in Fig. \ref{fig:backgroundSubtractedPSDplots} shows that the population is contained largely in the low-energy, low-PSP area of the region; this suggests a route by which one might improve on $\gamma$-ray discrimination, with marginal impact on neutron efficiency.
Optimization of accumulator configuration could be carried out in the spirit of the prescription of Ref. \cite{gatti1962}, and may lead to noncontiguous accumulator regions with rather different durations, such as those used in Ref. \cite{dolympia2012} to maximize neutron/$\gamma$ separation in CLYC.

\section{Acknowledgments}
Beneficial discussions with Michelle Faust, Lindsay Haselhorst, Dr. Jason Newby, and Dr. Nathaniel Bowden are very gratefully recognized.
This work was advanced tremendously by the nonpareil glass-machining skills of Peter Thelin.
The portion of this work conducted at LLNL was carried out under the auspices of the United States Department of Energy under contract number DE-AC52-07NA27344. 
Other portions of this work were supported under contract DE-FG02-97ER4104.
LLNL-JRNL-662220.

\bibliography{compositeNeutronDetector-GCRich-LLNL2014}

\end{document}